\documentclass[12pt]{article}

\usepackage{amsmath}
\usepackage{graphicx}
\usepackage{booktabs}
\usepackage{amsfonts}
\usepackage{hyperref}
\usepackage{amssymb}

\usepackage{amssymb}

\renewcommand{\theequation}
{\arabic{section}.\arabic{equation}}

\makeatletter
\def\eqnarray{ \stepcounter{equation} \let\@currentlabel=\theequation
 \global\@eqnswtrue
 \global\@eqcnt\z@
 \tabskip\@centering
 \let\\=\@eqncr
 $$\halign to \displaywidth\bgroup\@eqnsel\hskip\@centering
 $\displaystyle\tabskip\z@{##}$&\global\@eqcnt\@ne
 \hfil$\displaystyle{{}##{}}$\hfil
 &\global\@eqcnt\tw@$\displaystyle\tabskip\z@{##}$\hfil
 \tabskip\@centering&\llap{##}\tabskip\z@\cr}
\makeatother

\makeatletter
\def\@arrayacol{\edef\@preamble{\@preamble \hskip .5\arraycolsep}}
\def\array{\let\@acol\@arrayacol \let\@classz\@arrayclassz
\let\@classiv\@arrayclassiv \let\\\@arraycr\def\@halignto{}\@tabarray}
\makeatother

\makeatletter
\newcounter{subeqncnt}
\def\thesubeqncnt{\alph{subeqncnt}}
\def\subequations{\begingroup%
   \stepcounter{equation}\edef\@tempa{\theequation}%
   \let\c@equation\c@subeqncnt\c@subeqncnt\z@
   \edef\theequation{\@tempa\noexpand\thesubeqncnt}}

\makeatother

\newcommand{\be}{\begin{equation}}
\newcommand{\ee}{\end{equation}}

\newcommand{\bea}{\begin{eqnarray}}
\newcommand{\eea}{\end{eqnarray}}
\newcommand{\nn}{\nonumber}

\def\Det{{\rm Det}}

\begin{document}

\setlength{\baselineskip}{7mm}
\begin{titlepage}
 \begin{flushright}
{\tt NRCPS-HE-44-2018}
\end{flushright}

\begin{center}
{\Large ~\\{\it   Spectral Test of the MIXMAX Random Number Generators
\vspace{1cm}

}

}

\vspace{1cm}

 {\sl  Narek Martirosyan\footnote{${}$ On a leave of absence  from the 
 A.I. Alikhanyan National Science
Laboratory,Yerevan, 0036, Armenia}, Konstantin Savvidy  and 
George Savvidy

 \bigskip
 \centerline{${}$ \sl Institute of Nuclear and Particle Physics}
\centerline{${}$ \sl Demokritos National Research Center, Ag. Paraskevi,  Athens, Greece}
\bigskip
}

\end{center}
\vspace{30pt}

\centerline{{\bf Abstract}}

An important statistical test on the pseudo-random number generators is called the
spectral test. The test is aimed at answering the question of distribution of the generated 
pseudo-random vectors in dimensions $d$ that are larger than the genuine dimension of a generator $N$. 
In particular, the default  MIXMAX generators have various dimensions: $N=8,17,240$ and higher.  
Therefore the spectral test is important to perform in dimensions $d > 8$ for $N=8$ generator, 
$d> 17$ for $N=17$ and $d> 240$ for $N=240$ generator.  These tests have been performed 
by L'Ecuyer and collaborators. When $d > N$ the vectors of the generated 
numbers fall into the parallel hyperplanes and the distances between them can be larger than the 
genuine resolving power of the MIXMAX generators, which is $ l=2^{-61}$. The aim of this article is
to further study the spectral properties of the MIXMAX generators, to investigate the dependence 
of the spectral properties of the MIXMAX generators as a function of their  internal 
parameters and in particular their dependence on the parameter $m$. We found that the best
spectral properties are realised  when $m$ is between  $2^{24}$ and $2^{36}$, a range which is inclusive of the value of the $N=17$ generator. We also provide the alternative parameters for the generators, $N=8$ and $N=240$ with $m$ in this optimised range.

\vspace{12pt}

\noindent

\end{titlepage}

\pagestyle{plain}

\section{\it Introduction}

The MIXMAX generator of pseudo-random numbers \cite{savvidy_1986,konstantin_2015,konstantin_2016,savvidy_2015,Savvidy:2018ygo} demonstrates excellent statistical properties \cite{CLHEP,Ivantchenko,root} and is based on Anosov-Kolmogorov C-K systems  \cite{anosov_1967,kolmo,kolmo1}.   This innovative class of random number generators (RNGs) was proposed earlier by the members of the MIXMAX network and it relies on the fundamental discoveries and results of ergodic theory \cite{sinai3,rokhlin,rokhlin2}. 
The MIXMAX generator represents a hyperbolic dynamical system on $N$-dimensional torus.  Let $M $ be the $N$-dimensional torus and $\textbf{u} = (u_1, \hdots, u_N)^T$ be a point in $M$, $T$ denotes transpose. The mapping 
\be\label{Anosov}
\textbf{u} \mapsto A \textbf{u}~\mod ~ 1
\ee
realizes an Anosov  automorphism of $M$ and represents a Kolmogorov K-system  if $A$ is an integer matrix and satisfies the following two conditions  \cite{anosov_1967,savvidy_1986,konstantin_2015}:
\begin{eqnarray}
\label{ergodicity}
1)~\Det  A=  {\lambda_1}\,{\lambda_2}...{\lambda_N}=1,~~~~~
2)~~\vert {\lambda_i} \vert \neq 1, ~~~\forall ~~i .~~~~~~
\end{eqnarray}
The conditions (\ref{ergodicity}) ensure that such an automorphism represents  an Anosov C-system \cite{anosov_1967} and a Kolmogorov K-system \cite{kolmo,kolmo1,sinai3,rokhlin,rokhlin2},  exhibits a mixing  of  all orders and has nonzero Kolmogorov entropy.  In \cite{savvidy_1986} the authors suggested to use exponentially unstable trajectories  of the Anosov-Kolmogorov C-K systems to generate pseudo-random  numbers of high quality.

The trajectories of a C-system can be periodic and non-periodic.   All trajectories which start from vectors $\textbf{u}_0 \in M$ with rational coordinates, and only they, are periodic \cite{savvidy_2015, anosov_1967}. The rational numbers are everywhere dense on the phase space of a 
torus and the periodic trajectories of the C-systems follow the same pattern 
and are everywhere dense  \cite{anosov_1967}, 
like rational numbers on a real line.  

Thus if the initial vector $\textbf{u}_0 \in M$ has rational coordinates $u_i = a_i/p$, then the mapping (\ref{Anosov})  generates exponentially unstable periodic trajectories  on a sub-lattice $(a_1/p,...,a_N/p)$ of the Anosov-Kolmogorov C-K systems  and they  are used to generate  pseudo-random  numbers \cite{konstantin_2015,konstantin_2016}.    The high resolution  MIXMAX generators in the present form are  realised on a sub-lattice for which $p = 2^{61}-1$ \cite{konstantin_2015}. For a two-parameter family of C-system operators $A(N,s)$ which are
parametrised by the integers $N$ and $s$, the matrix has the following form \cite{konstantin_2015}:
\be
\label{eq:matrix}
A(N,s) =
   \begin{pmatrix}
      1 & 1 & 1 & 1 & ... &1& 1 \\
      1 & 2 & 1 & 1 & ... &1& 1 \\
      1 & 3{+}s & 2 & 1 & ... &1& 1 \\
      1 & 4 & 3 & 2 &   ... &1& 1 \\
      &&&...&&&\\
      1 & N & N{-}1 &  ~N{-}2 & ... & 3 & 2
   \end{pmatrix} 
\ee
and in \cite{konstantin_2016} there was introduced a three-parameter family of  operators $A(N,s,m)$: 
\be
\label{eq:matrix1}
A(N,s,m) =
   \begin{pmatrix}
      1 & 1 & 1 & 1 & ... &1& 1 \\
      1 & 2 & 1 & 1 & ... &1& 1 \\
      1 & m+2+s & 2 & 1 & ... &1& 1 \\
      1 & 2m+2 & m+2 & 2 &   ... &1& 1 \\
      1 & 3m+2 & 2m+2 & m+2 &   ... &1& 1 \\
      &&&...&&&\\
      1 & (N-2)m+2 & (N-3)m+2 &  (N-4)m+2 & ... & m+2 & 2
   \end{pmatrix},
\ee
which has larger entropy. In order to define the  MIXMAX generator one should find those values 
of $N$ and $s$ for which all the necessary conditions are fulfilled. 
For the  three-parameter family $A(N,s,m)$ of the MIXMAX generators the
optimal  values of the parameters are provided in Table \ref{tbl:largeM}.
The efficient implementation in software can be achieved for some particularly
convenient values of $m$ of the form $m=2^k+1$ \cite{konstantin_2016}.
Inspecting the data in the Table \ref{tbl:largeM} one can get convinced that
the system with $N=240,  s=487013230256099140$ and $m=2^{51}+1$
has the best stochastic properties within the $A(N,s,m)$ family of
operators in dimensions $d \leq 240$.   The ability to increase further the dimension $N$, the entropy 
 and the periods of the MIXMAX generators (\ref{eq:matrix1}) is a priceless advantage of the MIXMAX family  of RNGs allowing  them to pass strong statistical tests and in particular the BigCrush suite\cite{pierr}.  
\begin{table}[htbp]
   \centering
   \begin{tabular}{@{} lcccccl @{}}
      \toprule
      Size & Magic &Magic&Entropy& Log of the period  q\\
      N    & $m$ & s& &$\approx \log_{10} (q)$  \\
      \midrule
     8    & $m=2^{53}+1$&s=0 &220.4& 129\\
      17  & $m=2^{36}+1$&s=0& 374.3&294\\       
       240  & $m=2^{51}+1$& s=487013230256099140&8679.2& 4389\\

    \bottomrule
   \end{tabular}
   \caption{Table of three-parameter MIXMAX generators $A(N,s,m)$ in (\ref{eq:matrix1}). These generators have an advantage of having a very high quality sequence for moderate and small $N$. In particular, the smallest generator we tested, $N=8$, passes all the tests in the BigCrush suite
   \cite{pierr}.   }
   \label{tbl:largeM}
\end{table}
The behaviour of the correlation functions of the MIXMAX generators have been performed 
in \cite{Savvidy:2018ygo} and the results are presented in Table \ref{tbl:largeS}. 

The additional strong test of RNGs is provided by the so called 
spectral test \cite{marsaglia_1968,beyer_1971,dieter_1974,knuth_1981,afflerbach_1986}. {\it This test is aimed at answering the question of distribution of the generated random vectors in dimensions $d$ that are larger than the genuine dimension of the generator $N$}. 
In particular, for the generators presented in Tables \ref{tbl:largeM}, \ref{tbl:largeS} these dimensions are $d > 8$ for $N=8$ generator, $d> 17$ for $N=17$ and $d> 240$ for $N=240$ generator.  When $d > N$ the distribution of  the generated random vectors falls into the parallel hyperplanes and the distances between them can be larger than {\it the 
genuine resolving power of the MIXMAX generators, which is $l= {1\over p} \approx 10^{-18}$ in all dimensions $d \leq N$. }
\begin{table}[htbp]
   \centering
     \begin{tabular}{@{} cllccc @{}} 
      \toprule
      Dimension &~ Entropy & Decorrelation Time &  Iteration Time  & Relaxation Time                                 &Period  q\\
      N     &~~ $~h(T)$   &~~~~ $\tau_0 = {1\over h(T) 2N }$ & t & $\tau ={1\over h(T) \ln {1\over \delta v_0}}$ &  $  \log_{10} (q)$  \\ 
      \midrule
       8     &~~~~~220   & $~~~~~0.00028$     &1&1.54   & ${129}$ \\
       17   &~~~~~374       &~~~~~0.000079       &1&1.92   &294 \\
       240  &~~~~~8679   &~~~~~0.00000024    &1&1.17   &~4389\\
      \bottomrule
   \end{tabular}
   \caption{The MIXMAX parameters  $\tau_0~ < t ~< \tau $.  The iteration time $t$ is normalised to 1. The MIXMAX is a genuine 61 bit generator  on  Galois field GF[p], with Mersenne prime number $p=2^{61}-1$.  The initial volume element is $\delta v_0= 2^{-61 N}$.   
}
   \label{tbl:largeS}
\end{table}

It is the aim of this article to study the spectral properties 
of the MIXMAX generators. The initial investigation of the spectral index was performed in the article \cite{spyros_2018}. The authors calculated the spectral index for the operators $A(N,0)$ of the form (\ref{eq:matrix}) in dimensions $d=2N$ and have found that it is equal to $1/\sqrt{3}$.   Because this number is of order one, they concluded  that the distribution of the vectors generated by the operator $A(N,0)$   in $2N$ dimensions was not good.  This was one of the reasons to introduce the three parameter family of operators $A(N,s,m)$ given in (\ref{eq:matrix1}).  In \cite{lecuyer_2017} L'Ecuyer and collaborators  have studied the lattice structure of the general MIXMAX generator $A(N,s,m)$ in dimensions $d > N $ for the generators presented in Table \ref{tbl:largeM}. They confirm the result found in \cite{spyros_2018} for the $A(N,0)$ operator and have calculated the spectral index for the generators $N=8$ and $N=240$ with parameter $m$ presented in the Table \ref{tbl:largeM} and have found that the spectral indices are in the range $l \approx 10^{-3} - 10^{-4}$.  We would like to further investigate the dependence of the spectral properties of the MIXMAX generators as a function of their  parameters and in particular their dependence on the parameter $m$ given in Tables \ref{tbl:largeM},\ref{tbl:largeQ}. As we shall prove here the spectral index essentially depends on $m$. Our results are presented in Table \ref{tbl:largeQ}.   We found that the best spectral properties are realised  when $m$ is between  $2^{24}$ and $2^{36}$. In that case the spectral index is in the range $l \approx 10^{-8}-10^{-10}$ and is in the range which is inclusive of the value of the $N=17$ generator. 

 The conclusions which were drawn in article \cite{lecuyer_2017} by L'Ecuyer and collaborators  should not be overestimated,  because they are limited to the specific parameter values presented in Table \ref{tbl:largeM}. As we demonstrated in this paper they can be circumvented using the unprecedented freedom provided by the large parameter space of the C-K systems \cite{savvidy_1986,konstantin_2015,konstantin_2016,savvidy_2015} and as it follows from the results presented in Table \ref{tbl:largeQ}.

In the next sections we shall analyse the lattice structure of the MIXMAX generator in dimensions $d > N$. In the second section we shall review the basics of the lattice theory useful for performing the spectral test,  shall describe the dual lattices, the corresponding shortest vectors and the distances between adjacent hyperplanes.  The spectral test of random number generators is presented in the third section.  In the forth section the lattice structure is analysed when the skipping of some coordinates is performed.  In the fifth section we  analyse the spectral properties of the MIXMAX generators as a function of their internal parameters and we found that the best possible spectral index is of order $l_2=10^{-8}$. In the sixth section we demonstrate that the integration of $k$-time differentiable functions in $d > N$ is bounded by $l^k = 10^{-8 k}$.

\section{\it Dual Lattice and Distances Between Adjacent Hyperplanes}

For $m$ linearly independent basis vectors 
$\textbf{v}_1, \textbf{v}_2 \hdots, \textbf{v}_m \in R^d$, 
a lattice is the set of all points (vectors) which are constructed  by linear integer combinations of them: 
$$
\wedge =\left\{ \textbf{g} \in R^d ~~|~~ \textbf{g} = \sum_{i=1}^{m} z_i \textbf{v}_i, ~~z_i \in Z \right\},
$$ 
here $m$ is called the rank of the lattice and $d$ is its dimension. The lattice is said to have full rank if $m = d$. Let us define the matrix $\textbf{V} \in R^{d \times m}$ with the basis vectors $\textbf{v}_i$ as columns:
\be
\label{eq:spect}
\textbf{V} = \left\{\textbf{v}_1, \textbf{v}_2 \hdots, \textbf{v}_m\right\}=
   \begin{pmatrix}
      v^{(1)}_1 &v^{(1)}_2 & ... & ... & ... &v^{(1)}_{m-1}& v^{(1)}_m \\
      v^{(2)}_1 & v^{(2)}_2& ... & ... & ... &v^{(2)}_{m-1}& v^{(2)}_m \\
      ... & ... & ... & ... & ... &...& ... \\
      ... & ... & ... & ... &   ... &...& ... \\
      &&&...&&&\\
      v^{(d)}_1 & v^{(d)}_2 & ... & ... & ... & v^{(d)}_{m-1} & v^{(d)}_{m}
   \end{pmatrix},
\ee
which is called the {\it basis matrix}. Using $\textbf{V}$ one can equivalently define the lattice as 
$$
\wedge = \left\{\textbf{V} \cdot \textbf{z}~~ |~~ \textbf{z} \in Z^m \right\}.
$$ 
The same lattice points can be generated by many basis.  Any two basis $\textbf{V}_1, \textbf{V}_2 \in R^{d \times m}$ generate the identical lattice if $\textbf{V}_1 = \textbf{V}_2 \textbf{U}$, where $\textbf{U}\in Z^{m\times m}$ is a unimodular matrix, $det (\textbf{U}) = \pm 1 $. 

The important feature of a given lattice is its {\it fundamental parallelepiped} defined as the set of points 
$$
\mathcal{P} (\textbf{V})=\left\{\textbf{V} \cdot \textbf{t} ~~|~~ \textbf{t} \in [0,1)^m \right\}.
$$
The parallelepiped $\mathcal{P} (\textbf{V})$ is not uniquely  defined because its definition depends on the 
choice of the basis $\textbf{V}$, but its volume 
$$
\textrm{Vol}(\mathcal{P} (\textbf{V})) = \sqrt{\det(\textbf{V}^T \textbf{V})} \equiv \det(\wedge)
$$
is invariant under unimodular changes of the basis $\textbf{V} \rightarrow  \textbf{V}  \textbf{U}$.  If a lattice is of the full rank then $ \det(\wedge) = | \det(\textbf{V})|$. It is obvious that lattices with smaller determinants are denser 
populated by points.

The dual or reciprocal lattice $\wedge^{\star}$  is defined as the set of vectors $\textbf{y} \in R^{*d}$ which have integer scalar product with any of the vector $\textbf{g} \in R^d$ in $\wedge$:
\begin{eqnarray}
\label{dual_lattice}
\wedge^{\star} = \left\{\textbf{y} \in R^{*d} ~~|~~  \textbf{y} \cdot \textbf{g} = n \in Z, ~~\textbf{g} \in \wedge \right\} .
\end{eqnarray}
The dual lattice $\wedge^{\star}$ is generated by the basis matrix  \cite{micciancio_2012}
$$
\textbf{V}^{\star} = \textbf{V} (\textbf{V}^T \textbf{V})^{-1} \in R^{*d \times m},
$$
hence 
$$
\det (\wedge^{\star}) = {1\over \det (\wedge)}.
$$ 
If original lattice $\wedge$ is of a full rank, then $\textbf{V}$ is a square matrix and $\textbf{V}^{\star}=(\textbf{V}^T)^{-1}$. 

It follows from (\ref{dual_lattice}) that each dual vector $\textbf{y}$ defines a set of equally spaced parallel hyperplanes of the original lattice which are orthogonal to $\textbf{y}$. The distance between adjacent hyperplanes is 
\be\label{distance}
l = {1 \over  \vert \textbf{y} \vert },
\ee
where $\vert \textbf{y} \vert$ is the Euclidean length. In order to get some intuition, it is useful to consider two-dimensional case when original lattice is divided into parallel lines.
There are many ways of partitioning the lattice into hyperplanes, each of which corresponds to a different dual vector $\textbf{y}$. The shorter the dual vector $\textbf{y}$ is, the bigger is the distance between hyperplanes.

In physics, dual lattices play a central role in the theory of diffraction to describe the interaction of electromagnetic waves with crystals. Here the dual lattice is a lattice in momentum space, called reciprocal lattice. For example, when X-rays are scattered from a crystal, peaks in the intensity (constructive interference) of scattered radiation occurs if the momentum difference $\triangle \textbf{k}$ between incoming and diffracted X-rays
satisfy Laue condition:
\be
\triangle \textbf{k} \cdot \textbf{x} = n, ~n \in Z,~~~ \text{or equivalently}~~~ e^{2 \pi \, i \triangle \textbf{k} \cdot \textbf{x}} = 1. 
\ee
Here $\textbf{x}$ are the position vectors of atoms and $\textbf{k} =  \hat{\textbf{k}}/ \lambda $ is a wavevector with wavelength $\lambda$. It is assumed that scattering is elastic. Denoting $\textbf{y}=\triangle \textbf{k}$, we see that $\textbf{y} \in \wedge^{\star}$. Now all points satisfying Laue condition for a given $\textbf{y}$ lie on parallel planes separated by the distance $l = 1/\vert \textbf{y} \vert$, hence the scattering can be viewed as a reflection from a set of parallel planes (orthogonal to $\textbf{y}$) at some angle, called the Bragg angle $\theta$. In this way it can be shown that the Laue condition can be reduced to the Bragg condition $2 l \sin{\theta} = n \lambda$. Thus dual vectors describe the planes from which the diffraction pattern occurs.

\section{\it Spectral test of Random Number Generators}

The MIXMAX recurrence is given by (\ref{Anosov}), i.e. at each step $N$-dimensional vector $\textbf{u}_i = (u_{i~1}, \hdots, u_{i~N})$  is produced. 
Now consider the vectors 
\begin{eqnarray}
\label{vectors}
\textbf{g}_i = (\textbf{u}_i, \textbf{u}_{i+1}, \hdots, \textbf{u}_{i+r-1})^T \in [0,1)^{r N},~~i=0,...,p^N-1,
\end{eqnarray}
i.e. the combination of $r$ successive outputs of the generator. The set of all these vectors (or points) 
on the unit hypercube $[0, 1)^{r N}$ form a lattice structure described by the following basis matrix \cite{afflerbach_1988}:
\begin{equation}
\label{lcmg_basis}
\textbf{V}=    \begin{pmatrix}
  I/p & \textbf{0} & \cdots   & \textbf{0} \\
  A/p & I & \cdots & \textbf{0} \\
    \vdots  & \vdots  & \ddots & \vdots  \\
  A^{r-1}/p & \textbf{0} & \cdots & I 
 \end{pmatrix},
\end{equation}
where $I$ is an identity matrix of size $N$, and $\textbf{0}$ is an $N \times N$ 
matrix consisting of all zeros. Note that the lattice is of a full rank, $\textbf{V} \in R^{r N\times r N}$.

It is desirable to have not only $N$-dimensional points,  but also $r N$-dimensional points to be uniformly distributed  in hypercube $[0, 1)^{r N}$ when $r> 1$, so that the spacing between parallel hyperplanes 
is as small as possible. As we have seen, the lattice can be covered by parallel hyperplanes in different ways, hence one has to consider all possible coverings and take maximum of all distances. The so-called spectral test \cite{coveyou_1967,knuth_1981} measures the maximum distance between hyperplanes. Finding the maximum distance $l_{r N}$ among all sets of parallel hyperplanes amounts to finding the shortest vector in the dual lattice. 

To perform the spectral test one should  construct the dual basis $\textbf{V}^{\star}$ of (\ref{lcmg_basis}) with $\textbf{V}^{\star}=(\textbf{V}^T)^{-1}$ and find the shortest vector in the dual space:
\begin{eqnarray}
l_{r } = \frac{1}{\lambda_{min}},~~~\text{where}~~~~ \lambda_{min} = 
\min\limits_{\textbf{y} \in \wedge^{\star} \setminus \left\{\textbf{0}\right\}} \vert \textbf{y} \vert.
\end{eqnarray}
The spectral test therefore reduces to the finding of the shortest vector  in the dual lattice.

The shortest vector problem (SVP) is one of the most important and well studied lattice problems with applications in number theory, cryptography  \emph{etc.} \cite{micciancio_2012,lenstra_1982}.
To find the shortest vector in a lattice it is reasonable to make the lattice basis as orthogonal as possible since orthogonal basis is obviously shorter. 
The LLL algorithm \cite{lenstra_1982} applying the Gram-Schmidt orthogonalization finds the so called L-reduced basis of a given basis. Then the shortest
vector in such a basis is used as an approximate solution to the SVP. The LLL algorithm runs in polynomial time 
and approximates the solution with the factor of $\gamma = 2^{\frac{d-1}{2}}$, i.e. the algorithm returns a vector of length less than or equal to $\gamma \lambda_{min}$. The LLL algorithm is implemented in many software packages, e.g. in Mathematica the function LatticeReduce[$\textbf{V}^*$] implements the LLL algorithm. 

Let us consider first the linear congruential generators (LCG). The generation of pseudo-random numbers by the linear congruential method $x_i = a x_{i-1}~(\mod ~p)$ is the most studied method. The normalised values $u_i = x_i/p$ form a sequence of uniformly distributed random numbers in $[0,1)$, $d =1$. It is well known that the set of $d$-dimensional points ($u_i, u_{i+1}, \hdots, u_{i+d-1}$) have a lattice structure in $d$-dimensional unit hypecube $[0,1)^d$ when $d > 1$ \cite{marsaglia_1968,beyer_1971,dieter_1974,knuth_1981,afflerbach_1986}. These lattice points lie on $(d-1)$-dimensional parallel hyperplanes. The big distance between hyperplanes implies that the unit hypercube is mainly empty, hence the points are not uniformly distributed in $[0,1)^d, ~d>1$. Thus the distance between adjacent hyperplanes can be used for the assessment of the quality of uniformity of $d$-dimensional points. Since the lattice points can be covered by parallel hyperplanes in various ways, all possible coverings have to be considered.  The spectral test determines the maximum distance between adjacent parallel hyperplanes over all possible coverings \cite{coveyou_1967}. The shorter the distance is, the better is the uniformity. The spectral test has been proved to be a very powerful theoretical test which can reveal the weaknesses of the RNGs \cite{knuth_1981}.

The multiple recursive \cite{knuth_1981} and matrix recursive generators (MCG) \cite{franklin_1964,tahmi_1982,niki_1984,niederreiter_1986,grothe_1987} are generalizations of LCGs. The MCG is given by the following recurrence:
\begin{eqnarray}
\label{lcmg}
\textbf{x}_i = A \textbf{x}_{i-1} ~\mod ~ p, 
\end{eqnarray}
where $A \in Z^{N \times N}$ is integer matrix with entries less than $p$, and $\textbf{x}_0 \in Z^{N}$ is a vector of integers less than $p$.  The spectral test can also be applied to MCGs since $d= r N $-dimensional points $\frac{1}{p}(\textbf{x}_i, \textbf{x}_{i+1}, \hdots, \textbf{x}_{i+r-1})$ formed by $d$ successive normalized outputs form a lattice structure in $r N$-dimensional unit hypercube \cite{afflerbach_1988}.

\section{\it Skipping Some Coordinates of Lattice Points}
\label{skip_coord}

Let $\wedge$ be a full rank lattice  generated by the basis  $\textbf{V} = \left\{\textbf{v}_1, \textbf{v}_2 \hdots, \textbf{v}_d\right\} \in R^{d \times d}$ (\ref{eq:spect}),  $\textbf{g}= \sum_{i=1}^{d} z_i \textbf{v}_i$. Suppose some coordinates of $d$-dimensional lattice points are skipped (or deleted). In particular, skipping the second coordinate of a 3-dimensional point $(g_1, g_2, g_3)$ is a map $(g_1, g_2, g_3) \mapsto (g_1, g_3)$.  This map can be represented by the following matrix:
\begin{eqnarray}
P =
   \begin{pmatrix}
      1~ & 0~ & 0 \\
      0~ & 0~ & 1 
   \end{pmatrix}, 
\end{eqnarray}
The action of $P$ on an arbitrary vector skips its second component:
\begin{eqnarray}
P \begin{pmatrix}
      g_1 \\
      g_2 \\
      g_3 \end{pmatrix} = \begin{pmatrix}
      g_1 \\
      g_3 \end{pmatrix}
\end{eqnarray}
By skipping coordinates in $d$-dimensional lattice points we obtain the new set of points whose lattice structure we would like to describe. In particular, to find the basis vectors that generate this lattice. 

First of all, skipping coordinates of lattice points is
equivalent to skipping corresponding components in the basis vectors, i.e. the map
$$
(g_{i}^{(1)}, g_{i}^{(2)}, \hdots, g_{i}^{(d)})  \mapsto (g_{i}^{(i_1)}, g_{i}^{(i_2)}, \hdots, g_{i}^{(i_s)}),
~~~~\forall i,
$$
is equivalent to
$$
(v^{(1)}_k, v^{(2)}_k, \hdots, v^{(d)}_k)  \mapsto (v^{(i_1)}_k, v^{(i_2)}_k, \hdots, v^{(i_s)}_k),
~~~~1 \leq k \leq d.
$$
This map can be represented by the following $P \in R^{s \times d}$ matrix,
\be
\label{map_matrix}
P = 
\begin{pmatrix}
      0  & ... & 0 & \delta_{i_1} & 0 & 0 & ... & 0 \\
      0  & ...  & 0 & 0 & \delta_{i_2} & 0 & ... & 0 \\
      ... & ... & ... & ... & ... &...& ... \\
      0  & ... & 0 & 0 & 0 & \delta_{i_s} & ... & 0
\end{pmatrix},
\ee
where $\delta_{i_s}=1$ indicates that the row has 1 in $i_s$-th entry and  0's elsewhere.

Now $\textbf{E} = P \textbf{V} \equiv \left\{ \textbf{e}_1, \textbf{e}_2, \hdots, \textbf{e}_d \right\}$ is the matrix whose columns 
are $s$-dimensional vectors with skipped components; $\textbf{e}_k = (v^{(i_1)}_k, v^{(i_2)}_k, \hdots, v^{(i_s)}_k)$.  

Let $\rho$ be the rank of the matrix $\textbf{E}$,  $\rho  \leq s $, so there are $\rho$ linearly independent 
vectors among $d$ vectors. Thus in order to describe the lattice structure obtained by skipping coordinates we have to choose $\rho$ linearly independent vectors from the set $\left\{\textbf{e}_1, \textbf{e}_2, \hdots, \textbf{e}_d \right\}$ so that all remaining vectors can be expressed as a linear integer combination of the chosen set. 

The null space of the matrix $\textbf{E}$ allows to find the relations between the vectors $\textbf{e}_1, \textbf{e}_2, \hdots, \textbf{e}_d$. Indeed, the null space $N(\textbf{E})$ of $\textbf{E}$ is a full set of solutions  $\textbf{x} = (x_1, \hdots, x_d)^T \in R^d$ of the equation
\be
\label{null}
\textbf{E} \textbf{x}=\textbf{0},
\ee
i.e.
\be
\label{nullspace}
N(\textbf{E})= \left\{{\textbf{x} \in R^d~~|~~ \textbf{E} \textbf{x}=\textbf{0}}\right\}.
\ee
Note that  $\textbf{E} \textbf{x}=\textbf{0}$
is equivalent to the equation  
\be\label{indexcal}
\sum_{k=1}^{d} x_k \textbf{e}_k = \textbf{0},
\ee
since the columns of the matrix $\textbf{E}$ are the vectors $\textbf{e}_k$. The solution vectors $\textbf{x}$ to (\ref{null}), (\ref{indexcal}) which are linearly independent form the basis of $N(\textbf{E})$, and there are ($d-\rho$) such solutions. The basis of $N(\textbf{E})$ allows to examine the relations (\ref{indexcal}) and find such $\rho$ linearly independent vectors which form a basis of the projected lattice. Having in hand the basis of the projected lattice one can analyze its spectral properties, and we shall use this method in the next sections.

\section{\it The Spectral Test of MIXMAX Generator}

If we consider the lattice structure of $r N$-dimensional points (\ref{vectors}) formed by $r$ consecutive outputs of the MIXMAX generator, $r > 1$, then independently of the parameters $N$ and $s$ of the operator $A(N,s)$ and of the three-parameter $A(N,s,m)$ family of operators, the shortest vector in the reduced dual lattice basis is $\sqrt{3}$, hence the spectral index is $l_{r N} = 1/\sqrt{3}$ \cite{spyros_2018,lecuyer_2017}. 
This lattice structure results from the relationships between certain coordinates of $r N$-dimensional points.  L'Ecuyer et al. \cite{lecuyer_2017} writing the dual basis explicitly found that the dual vector of length $\sqrt{3}$ corresponds to the following relationship between the second, the $(N+1)$-th and the $(N+2)$-th coordinates
of the (\ref{vectors}):
\begin{eqnarray}
\label{basic_relation}
g_{i}^{(2)} + g^{(N+1)}_i - g^{(N+2)}_i =  \begin{cases} 0 \\ 1  \end{cases}
\end{eqnarray}
Hence the relationship is absent if the first component of each generated MIXMAX vector is skipped. The following linear relation exists for the three-parameter operator $A(N,s,m)$ (\ref{eq:matrix}) when $j=4, ..., N-2 \mod N$:
\begin{eqnarray}
\label{advanced_relation}
g_{i}^{(j)} - 2 \, g^{(j+1)}_i + g^{(j+2)}_i -  g_{i}^{(j+N+2)}  - (m-1) \, g_{i}^{(j+N+1)} = 0 \mod 1. 
\end{eqnarray}
With $m=1$ it is also valid  for the two-parameter family $A(N,s)$  and implies the Proposition 3  in \cite{lecuyer_2017}.  
The relation (\ref{basic_relation}) is present for the three-parameter MIXMAX generators $A(N, s,m)$ (\ref{eq:matrix1}) hence skipping the first coordinate is also necessary there. The relation (\ref{advanced_relation}) implies a few other relationships  which do not involve the first two coordinates, see the Propositions 4, 5 and 6 in \cite{lecuyer_2017}. These propositions are employed for analysing the three-parameter generators of the dimensions $N=8$ and $N=240$, see examples 1 and 2 in \cite{lecuyer_2017}. From the relation (\ref{advanced_relation}) it follows that the spectral index is inversely proportional to $m$
\be
l_2 \approx {1\over m}.
\ee

Let us now consider the generator with the parameters $N = 17$, $m=2^{36}+1$ and $s = 0$. Taking $r=2$ and skipping only the first coordinate of each output gives the lattice with spectral index 
\be\label{mixmaxindex}
l_2 = 1.49 \cdot 10^{-8}  ~.
\ee
This is the main  advantage of the MIXMAX generators with parameter $m$ in the region between  $2^{24}$ and $2^{36}$. 
Taking $r=3$, one additional special relation exists which necessitates skipping the first two components of each output and gives the spectral index $l_3 = 0.00049$.
\begin{table}[htbp]
   \centering
   \begin{tabular}{@{} lccccclr @{}}
      \toprule
      N    & $m$ & s& Entropy&$   \log_{10} (q)$  & spectral index $l_2$  \\
      \midrule
         8    & $m=2^{36}+1$& s=0                 &149.7     & 129 &   $1.49 \cdot 10^{-8}$  \\   
          17    & $m=2^{36}+1$& s=0                 &374.3     & 294 &   $1.49 \cdot 10^{-8}$   \\   
              240  & $m=2^{32}+1$& s=271828282 & 5445.7&4389 &     $7.6 \cdot 10^{-10}$ \\
    \bottomrule
   \end{tabular}
   \caption{Table of the new three-parameter MIXMAX generators $A(N,s,m)$ in (\ref{eq:matrix1}).  
The parameters of the    $N=17$ generator are identical with the ones presented in Table \ref{tbl:largeM}. 
There is no need to change the parameters of the $N=17$ generator as far as its parameter $m$ is already in the preferred region $2^{24}- 2^{36}$. All generators defined on Tables \ref{tbl:largeM}  and \ref{tbl:largeQ} have passed the TestU01 and are the recommended generators for the Monte Carlo simulations.}
   \label{tbl:largeQ}
\end{table}
Based on the above, we have been able to construct the new improved values of the parameters for the generator with $N=8$:
$m=2^{36}+1, s=0$  and for $N=240$ we have $m=2^{32}+1, s=271828282$. These maximise the entropy and essentially improve  the spectral index. In order to maximise the entropy at the same time as the period and the spectral index it would be necessary to  move away from a near-power-of-two form of the $m$-parameter.

\section{ \it Practical Implications}

All Monte-Carlo simulations implicitly contain a function which represents the observable of interest to be averaged over random instances of some object under study. Thus, in almost all situations the problem 
can be mapped to integrating a function $f$ of $s$ real variables $f(u_1, ... , u_s)$ defined on the hypercube $R \in [0,1)^s$. That may require rescaling the variables to fit inside the unit cube and in the case that the instance of the object which is under study is generated using a variable but finite number of random variables then also padding may be required. We draw real variables $u_i$ from the RNG consecutively and calculated the average value of the
function f over the pseudo-random sequence. If the native dimension of the RNG is $N$, then for $s \le N$ the Monte-Carlo integration is guaranteed to converge to the correct result (within machine floating-point accuracy) and, moreover, the error is guaranteed to be normally distributed according to the theorem of Leonov. For $s > N$, the Monte-Carlo integration may not converge to the correct result if the function is of the form
\[ f(\mathbf{u})=\cos(2\pi \, \mathbf{y} \cdot \mathbf{u}), \]
where $\mathbf{y}$ is some vector of the dual lattice (\ref{dual_lattice}) \cite{coveyou_1967}. Worries arise if the $\mathbf{y}$-vector is short.  All such $\mathbf{y}$ can in principle be found explicitly and analysed. 

Let us consider few examples. Using the standard technique we can find the shortest vector in 121 dimensions in the dual lattice of the random number generator of L'Ecuyer \cite{lecuyer_1999} known as MRG32k3a. This generator has excellent properties in low dimensions. In higher dimensions one such short vector (among exponentially many of similar length) is
\begin{eqnarray}\label{they}
\mathbf{y} = &&( -1, -1, 0, 0, 0, 3, 0, 0, 1, -2, 2, 0, -1, -2, -1, 2, 0, 1, 0, 1, -1, 1, 0, \nn\\
                        &&-3, 0, -2, 0, 0, 1, 0, 0, 0, 1, 1, 0, 0, -2, 1, 0, -1,  1, 0, 0, 1, 0, 1, 1, 0, 1,\nn\\ 
                        &&0, 0, 1, 3, 0, 0, 0, 1, 0, 1, 0, -1, 0, -1, 1, 0, 0, -1, -1, 0, 0, 2, 2, 0, -2,-1,\nn\\ 
                        &&0, 0, 1, 1, 0, 1, 0, -1, 1, -2, 1, 2, 1, 1, -1, 0, 0, 0, 0, 0, 0, 1, 0, -2, -1, 0, \nn\\
                        &&0, 1, 1, 0, 0, 1, -1, 1, 0, 0, 2, 0, 1, -1, 0, 1, -1, 0, 0, -1 )^T
\label{eq}
\end{eqnarray}
 None of the components of this vector is larger than $3$ by absolute value and the length of the vector is approximately $\vert \mathbf{y} \vert =  11.3$ and the spectral index is 
 $l = {1 \over \vert \mathbf{y} \vert} = 0.088$.  Since there are only 65 non-zero components in the vector, an appropriate choice of a skipping schedule will result in the multi-set dimension equal to 65 under the definitions in \cite{lecuyer_2017} and earlier works. The
function  $f(\mathbf{u})=\cos(2\pi \, \mathbf{y} \cdot \mathbf{u}) $, where $\mathbf{y} $ is given by (\ref{they}), is  a smooth function without any sharp discontinuities or fast oscillations. The average of this function over the unit hypercube is zero. Nevertheless, as one can verify also with actual RNG software, the points generated by MRG32k3a all fall onto points where $f \approx 1$, and thus the Monte-Carlo answer fails to converge at all to the correct value. The situation is of course not specific to dimension 121, as all larger dimensions will have a spectral index worse than $0.1$. In dimensions larger than about 1000, one can conjecture that the MRG32k3a and most other MRG will have vectors in the reduced dual lattice mostly consisting of $0$, with a few values of $\pm 1$, and the properties will be equivalent to a lagged-Fibonacci generator.

Whether the value of the spectral index worse than $\approx 0.1$ should disqualify an RNG from being used in some high dimension is not at all clear from the theory, but some additional arguments can be considered. If the observable $f$ is $k$-times differentiable and supremum of all the $k$-th partial derivatives is bounded by a positive constant:
\be\label{bound}
 | f^{(k)} | < A ~, 
 \ee
then the corresponding Fourier series 
\[   f(\mathbf{u})= \sum_\mathbf{h \in \mathbb{Z}^s}  c_\mathbf{h} ~ \exp (i\,2\pi \, \mathbf{h} \cdot \mathbf{u})  
\]
will have the coefficients rapidly falling with the wavenumber:
\be\label{bound2}
 | c_\mathbf{h} | < \frac{A}{|\mathbf{h}|^k} ~
 \ee
Thus if the integer-component Fourier mode $\mathbf{h}$ coincides with some or another vector of the dual lattice $\mathbf{y}$ , and the corresponding coefficient $c_\mathbf{y}$ is not zero, then the value
$$\int f(\mathbf{u})   d \mathbf{u}=  c_0,$$  
of the overall integral will not be obtained. The Monte-Carlo sum will converge toward:
\be\label{error}
 \frac{1}{n} \sum^n_{i=1} f(\mathbf{g}_i)  \rightarrow c_0 + \sum_{\mathbf{y} \in \wedge^*} c_{\mathbf{y}} ~. 
 \ee
From the bound on the Fourier coefficients in (\ref{bound2}) one can see from a yet another point of view why a smaller spectral index $l \ll 1$ is generally beneficial: the value of the Fourier coefficient corresponding to some vector of the MIXMAX dual lattice $\mathbf{y}$ is suppressed  
$$ 
| c_\mathbf{y} | < \frac{A}{|\mathbf{y}|^k} ~<~ A ~l^k
$$  
as a positive power of the spectral index $l^k= 10^{-8 k} - 10^{-9 k}$. Thus, one may be able to bound the coefficient of the problematic modes to below the limit of machine precision. It is possible to use these estimates to prove that the sum of the error terms above (\ref{error}) will converge by absolute value, if $k>s+1$.  From this point of view, the MIXMAX generators having a value of the spectral index of $l_2 \approx 10^{-8}$ (\ref{mixmaxindex}) in dimensions $d  \leq 2\,N$ are not of particular concern.  
The alternative option to achieve an excellent spectral index in large dimensions is of course to use the MIXMAX generators with increasing $N$: the largest dimension of  MIXMAX generator which was provided in \cite{konstantin_2016} has $N=44851$,  thus for all dimensions $d \leq 44851$ it is 
\be
l={1\over p} \approx  10^{-18}.
\ee

In conclusion we would like to stress that the analysis of the spectral index in \cite{lecuyer_2017} is complete in all details of the results formulated in the Propositions 1-6, but there is a less appreciation of the mathematics of the C-K systems and of the new avenues that they open for applications.  {\it The conclusions which were drawn from Propositions 1-6 should not be overestimated,  because they are limited to the specific parameter values presented in Table \ref{tbl:largeM}. As we demonstrated in this paper they can be circumvented using the unprecedented freedom provided by the large parameter space of the C-K systems \cite{savvidy_1986,konstantin_2015,konstantin_2016,savvidy_2015} and as it follows from the results presented in Table \ref{tbl:largeQ}.}   The spectral indexes of the specific MIXMAX generators presented in Table \ref{tbl:largeQ} and in \cite{konstantin_2016} are vastly superior to most other random number generators in all dimensions considered above, and certainly of  generators proposed by the authors of \cite{lecuyer_2017}, as it follows from (\ref{they}) and the text afterword.

\section{\it Acknowledgement }
This work was supported in part by the European Union's Horizon 2020
research and innovation programme under the Marie Sk\'lodowska-Curie
Grant Agreement No 644121.

\section{\it Note Added }

The authors of the MIXMAX generators in [3] have introduced the  operator $A(N,s,m,b)$ depending on 
four parameters, where $N$ is the dimension of the matrix operator and $s,m,b$ are its {\it internal parameters}.  The initial vector $\textbf{u}_0 \in M$ has rational coordinates $u_i = a_i/p$, therefore the periodic trajectories  are defined on a sub-lattice $(a_1/p,...,a_N/p)$ of the Anosov-Kolmogorov C-K systems.    The high resolution  MIXMAX generators in the present form are  realised on a sub-lattice for which $p = 2^{61}-1$ and generate 61-bit random numbers. It is obvious that one can consider the periodic trajectories defined of the different sub-lattices   $(a_1/p_1,...,a_N/p_1)$ ,$(a_1/p_2,...,a_N/p_2),...$  for the same operator  $A(N,s,m,b)$.  Therefore it is incorrect to consider the $p$ as a parameter defining the operator $A(N,s,m,b)$,  it defines only a sub-class of periodic trajectories of the MIXMAX operator $A(N,s,m,b)$.  All generators defined on Tables \ref{tbl:largeM}  and \ref{tbl:largeQ} have passed the TestU01 and are the recommended generators for the Monte Carlo simulations.

\section{\it Appendix. Example of a Projected Lattice }

Let us illustrate the procedure described in section \ref{skip_coord} on the example of two-parameter MIXMAX generator. We take for simplicity $N=8$ and $r=2$. Hence (\ref{lcmg_basis}) reduces to
\begin{equation}
\textbf{V}=    \begin{pmatrix}
  I/p & \textbf{0} \\
  A/p & I 
 \end{pmatrix},
\end{equation}
where $I$ is an identity matrix of size $8$, $p=2^{61}-1$ and $A$ is given by (\ref{eq:matrix}) with $N=8$ and $s=0$.

Suppose the first six coordinates of each output are skipped, i.e. only the 7-th, the 8-th, the 15-th and the 16-th coordinates remained in 16-dimensional lattice points. The skipping of components in the basis vectors is represented by the following operator $P$ (\ref{map_matrix}):
\be
P = 
\begin{pmatrix}
      0 & 0 & 0 & 0 & 0 & 0 & 1 & 0 & 0 & 0 & 0 & 0 & 0 & 0 & 0 & 0 \\
      0 & 0 & 0 & 0 & 0 & 0 & 0 & 1 & 0 & 0 & 0 & 0 & 0 & 0 & 0 & 0 \\
0 & 0 & 0 & 0 & 0 & 0 & 0 & 0 & 0 & 0 & 0 & 0 & 0 & 0 & 1 & 0 \\
0 & 0 & 0 & 0 & 0 & 0 & 0 & 0 & 0 & 0 & 0 & 0 & 0 & 0 & 0 & 1 \\
\end{pmatrix}
\ee
Now $\textbf{E} = P \textbf{V} = \left\{ \textbf{e}_1, \textbf{e}_2, \hdots, \textbf{e}_{16} \right\}$ is the matrix with vectors $\textbf{e}_k = (v^{(7)}_k, v^{(8)}_k, v^{(15)}_k, v^{(16)}_k)$ as columns. The linearly independent solutions of (\ref{null}, \ref{indexcal}), i.e. the 
basis of $N(\textbf{E})$, correspond to the following relations: 
\begin{align}\label{nullequation}
    \textbf{e}_2 = \textbf{e}_3 = \textbf{e}_4=\textbf{e}_5=0 \nn \\
    x_{51} \textbf{e}_1 + x_{56} \textbf{e}_6 + x_{57} \textbf{e}_7 + x_{58} \textbf{e}_8 + \textbf{e}_9 = 0 \nn\\
     x_{61} \textbf{e}_1 + x_{66} \textbf{e}_6 + x_{67} \textbf{e}_7 + x_{68} \textbf{e}_8 + \textbf{e}_{10} = 0 \nn\\
     x_{71} \textbf{e}_1 + x_{76} \textbf{e}_6 + x_{77} \textbf{e}_7 + x_{78} \textbf{e}_8 + \textbf{e}_{11} = 0 \nn\\
     x_{81} \textbf{e}_1 + x_{86} \textbf{e}_6 + x_{87} \textbf{e}_7 + x_{88} \textbf{e}_8 + \textbf{e}_{12} = 0 \\
     x_{91} \textbf{e}_1 + x_{96} \textbf{e}_6 + x_{97} \textbf{e}_7 + x_{98} \textbf{e}_8 + \textbf{e}_{13} = 0 \nn\\
     x_{10\,1} ~\textbf{e}_1 + x_{10\,6}~ \textbf{e}_6 + x_{10\,7} ~\textbf{e}_7 + x_{10\,8}~ \textbf{e}_8 + \textbf{e}_{14} = 0\nn\\
     x_{11\,1}~ \textbf{e}_1 + x_{11\,6}~ \textbf{e}_6 + x_{11\,7}~ \textbf{e}_7 + x_{11\,8} ~\textbf{e}_8~ + \textbf{e}_{15} = 0\nn\\
     x_{12\,1}~ \textbf{e}_1 + x_{12\,6}~ \textbf{e}_6 + x_{12\,7}~ \textbf{e}_7 + x_{12\,8}~ \textbf{e}_8 + \textbf{e}_{16} = 0.\nn
\end{align}
In particular, the relation
\begin{eqnarray*}
 x_{51} \textbf{e}_1 + x_{56} \textbf{e}_6 + x_{57} \textbf{e}_7 + x_{58} \textbf{e}_8 + \textbf{e}_9 = 0
\end{eqnarray*}
corresponds to the solution of (\ref{null}) of the form
\begin{eqnarray*}
\textbf{x}_5= \begin{pmatrix}
x_{51} & 0 & 0 & 0 & 0 & x_{56} & x_{57} & x_{58} & 1 & 0 & 0 & 0 & 0 & 0 & 0 & 0
\end{pmatrix}^T.
\end{eqnarray*}
Note that in this case $\rho = 4$ and $d-\rho = 12$. 

Non-zero integers $x_{ij}$ are: $x_{51} = 401311745785675161,~ x_{56} = -159309952957684710,~ \\x_{57} = 40454363710136096, x_{58} = 69541385551308331,~ x_{61} = 216592440415012406,~ \\x_{66} = -31873135044349652, x_{67} = 46806701813380607,~x_{68} = 4234892068829674,~ 
\\x_{71} = -4234892068829674,~ x_{76} = -61071601413648983,~ x_{77}=5186463356723851, ~ \\x_{78}=24239184867643529,~ x_{81}=46806701813380607,~ x_{86}=-53159039916625118,~\\ x_{87}=207063933260145640,~ x_{88}=-25186463356723851,~ x_{91}=31873135044349652,~ \\x_{96}=95563682868985407,~ x_{97}=53159039916625119,~ x_{98}=-61071601413648983,~\\ x_{10~1}=-31594523724031910,~ x_{10~6}=151174502404406653,~ x_{10~7}=551984747813509878,~ \\x_{10~8}=-213583438155580796,~ x_{11~1}=-254372135450098182,~ x_{11~6}=700651748335056856,~\\ x_{11~7}=906044013673296045,~ x_{11~8}=-576836877585852439,~ x_{12~1} =-604865176409817255,~ \\x_{12~6}=1592820918256529364,~ x_{12~7}= 563352089682473740,~ x_{12~8}=-937248481548883117$.
 
As one can see from (\ref{nullequation}), the set of 4 vectors $\left\{ \textbf{e}_1, \textbf{e}_6, \textbf{e}_7, \textbf{e}_8\right\}$ form the basis of projected lattice by which all other vectors $\textbf{e}_k$ ($k\neq 1,6,7,8$) can be expressed by  linear integer combinations. Having in hand the basis of the projected lattice one can analyse its spectral properties.

\vfill

\end{document}